\documentclass[aps,prb,twocolumn,longbibliography,groupedaddress,showpacs,floatfix,superscriptaddress]{revtex4-1}
\usepackage{epsfig}
\usepackage{amsmath,amssymb}
\usepackage{graphicx}
\usepackage[dvipsnames,usenames]{color}
\usepackage[normalem]{ulem}
\emergencystretch=\maxdimen
\begin{document}

\title{The two-dimensional $t$-$t'$ Holstein model}
\author{Maykon V. Ara\'ujo} 
\affiliation{Departamento de F\'isica, Universidade Federal do Piau\'i,
64049-550 Teresina PI, Brazil}
\author{Jos\'e P. de Lima} 
\affiliation{Departamento de F\'isica, Universidade Federal do Piau\'i,
64049-550 Teresina PI, Brazil}
\author{Sandro Sorella}
\affiliation{International School for Advanced Studies (SISSA),
Via Bonomea 265, 34136, Trieste, Italy}
\author{Natanael C. Costa}
\affiliation{International School for Advanced Studies (SISSA),
Via Bonomea 265, 34136, Trieste, Italy}
\affiliation{Instituto de F\'isica, Universidade Federal do Rio de
Janeiro Cx.P. 68.528, 21941-972 Rio de Janeiro RJ, Brazil}

\begin{abstract}
The competition and interplay between charge-density wave and superconductivity have become a central subject for quasi-2D compounds.
Some of these materials, such as the transition-metal dichalcogenides, exhibit strong electron-phonon coupling, an interaction that may favor both phases, depending on the external parameters, such as hydrostatic pressure.
In view of this, here we analyze the single-band $t$-$t^{\prime}$ Holstein model in the square lattice, adding a next-nearest neighbor hopping $t^{\prime}$ in order to play the role of the external pressure.
To this end, we perform unbiased quantum Monte Carlo simulations with an efficient inversion sampling technique appropriately devised for this model. Such a methodology drastically reduces the autocorrelation time, and increases the efficiency of the Monte Carlo approach. By investigating the charge-charge correlation functions, we obtain the behavior of the critical temperature as a function of $t^{\prime}$, and from compressibility analysis, we show that a first-order metal-to-insulator phase transition occurs. We also provide a low-temperature phase diagram for the model.
\end{abstract}


\maketitle

\section{Introduction}

The emergence of charge-density wave (CDW) and superconductivity (SC) in transition-metal dichalcogenides (TMDs) has been a matter of intense debate over the past decades\cite{Zhu15,Zhu17,Manzeli17}.
The ability to tune these phases is the key feature to better understanding their nature, e.g., as in a recent gate-induced experiment on few layers of the 1T-TiSe$_{2}$\,\cite{Li16}, which remarkably showed the occurrence of a Kosterlitz-Thouless phase transition.
In view of this, a great experimental effort has been done to characterize these compounds by different external parameters: by changing the number of (chalcogen-metal-chalcogen) layers, chemical doping/intercalation, strain or hydrostatic pressure\,\cite{Chen16,Wagner08,Li16b,Li17,Cho18,Bu19}.
Interestingly, the suppression of the CDW phase, for most of the cases, is followed by the appearance of a superconducting dome around the critical point, resembling the phase diagrams of heavy fermion materials or doped cuprates\,\cite{Chatterjee15}.
Thus, investigating the features of the CDW phase, and how SC emerges, may lead to insights into the basic properties of TMDs and other compounds.

From a theoretical point of view, it is important to understand the most fundamental features of these phases, i.e., how external parameters affect charge and pairing correlations.
It is also worth noticing that, different from chemical doping/intercalation, which could lead to electronic doping or disorder effects, the application of pressure leaves the material clean, avoiding further complexities to theoretical approaches.
Thus, we examine such an interplay in the physical background of applied hydrostatic pressure, using effective Hamiltonians.
Finally, since the electron-phonon coupling plays a crucial role in the emergence of long-range order in TMDs, such interactions are indispensable to our model formulation.

Given this, we analyze the properties of the single-band $t$-$t^{\prime}$ Holstein model in the square lattice\,\cite{Holstein59}.
This effective Hamiltonian has been extensively adopted for 1D, 2D, and 3D geometries as a `standard model' for compounds exhibiting phonon-induced features.
In particular, its properties have been explored in many different aspects, e.g.~by examining the effects of anharmonicity, strain, disorder, or electronic doping\,\cite{Vekic92,Freericks96b,Jeckelmann99,Hohenadler04,Tozer14,Chen19,CohenStead19,Li18,Costa18,Costa20,Feng20,Costa21,Zhang19,Xiao21b,Dee19,Dee20,Bradley21,Paleari21,Cohen-Stead20,Stolpp20}.  
Here, instead, we investigate how the overlap between farther orbitals affects the leading CDW order.
Within our approach, the next-nearest neighbor (NNN) electron hopping, $t^{\prime}$, plays the role of hydrostatic pressure. 
In addition, this model allows us to study features of charge frustration, an issue less explored in the literature\,\cite{Li18}.
To this end, we develop and use a state-of-the-art quantum Monte Carlo (QMC) approach, which provides us the correlation functions and, therefore, the critical points of the model. 

The organization of this paper is as follows: In Sec.\,\ref{Sec:ModelandMethod} we define the Hamiltonian, and review the methodology, defining the observables of interest.
Section \ref{Results} presents our results, emphasizing the effects of pressure on the critical CDW temperatures.
Our conclusions and further comments are in Sec.\,\ref{Conclusions}.

\section{Model and Methodology} \label{Sec:ModelandMethod}


The $t$-$t'$ Holstein Hamiltonian \cite{Holstein59,Vekic92} reads
\begin{align} \label{eq:Holst_hamil}
\nonumber \mathcal{H} = &
-t \sum_{\langle \mathbf{i}, \mathbf{j} \rangle, \sigma} 
\big(d^{\dagger}_{\mathbf{i} \sigma} d^{\phantom{\dagger}}_{\mathbf{j} \sigma} + {\rm h.c.} \big)
-t^{\prime} \sum_{\langle \langle \mathbf{i}, \mathbf{j} \rangle \rangle, \sigma} 
\big(d^{\dagger}_{\mathbf{i} \sigma} d^{\phantom{\dagger}}_{\mathbf{j} \sigma} + {\rm h.c.} \big)
\\
&
- \mu \sum_{\mathbf{i}, \sigma} n^{\phantom{\dagger}}_{\mathbf{i}, \sigma}
+ \sum_{ \mathbf{i} }
\bigg( \frac{\hat{P}^{2}_{\mathbf{i}}}{2 M} + \frac{M \omega^{2}_{0} }{2} \hat{X}^{2}_{\mathbf{i}}\bigg)
- g \sum_{\mathbf{i}, \sigma} n_{\mathbf{i}\sigma} \hat{X}_{\mathbf{i}} ~,
\end{align}
where the sums over $\mathbf{i}$ run in a two-dimensional square lattice, with $\langle \mathbf{i}, \mathbf{j} \rangle$ and $\langle \langle \mathbf{i}, \mathbf{j} \rangle \rangle$ denoting nearest and next-nearest neighbors sites, respectvely.
We work within a second quantization formalism, in which $d^{\dagger}_{\mathbf{i} \sigma}$ and $d_{\mathbf{i} \sigma}$ are ordinary creation and annihilation operators of electrons with spin $\sigma$ at a given site $\mathbf{i}$, and $n_{\mathbf{i}, \sigma}\equiv d^{\dagger}_{\mathbf{i} \sigma} d_{\mathbf{i} \sigma}$ are number operators.
The bare phonon modes are added by local harmonic oscillators of frequency $\omega_{0}$, with $\hat{P}_{\mathbf{i}}$ and $\hat{X}_{\mathbf{i}}$ being conjugate momentum and position operators, respectively.
The first two terms on the right-hand side of Eq.\,\eqref{eq:Holst_hamil} correspond to the electron kinetic energy operators, with $t$ ($t^{\prime}$) being the hopping integral of nearest (next-nearest) neighbor orbitals.
The third one denotes the chemical potential $\mu$, while the fourth term describes the dispersionless bare phonon modes.
Finally, the electron-phonon interaction is given in the last term.
Hereafter, for convenience, we define the masses of the harmonic oscillators ($M$) and the lattice spacing ($a$) as unity, while setting $t$ as the scale of energy.

To the best of our knowledge, the first and single attempt to study the model presented in Eq.\,\eqref{eq:Holst_hamil} by QMC approaches was given in Ref.\,\onlinecite{Vekic92}.
However, besides the small lattice sizes achieved, the authors were concerned about features away from half-filling.
Here, on the other hand, we are interested to investigate how $t^{\prime}$ changes the charge and pairing correlation functions at the \textit{half-filling}, i.e., for $\langle n_{\mathbf{i} \sigma} \rangle = 1/2$.
It requires the adjustment of $\mu$ for each value of temperature and $t^{\prime}/t$.
Indeed, this particular filling is of great interest to our work: at $t^{\prime}=0$, the noninteracting Hamiltonian exhibits Fermi surface nesting (FSN), which leads to CDW for any finite electron-phonon coupling at the square lattice\,\cite{Costa20}.
Then, the addition of a NNN hopping would provide insights on the effects of pressure for such a CDW phase. 

Before proceeding, it is important defining the external parameters, and the physical quantities of interest.
Following previous works, we define the dimensionless eletron-phonon coupling as $ \lambda_{D} = g^2/(\omega_{0}^2 W) $, with $W=8t$ being the noninteracting electronic bandwidth (for $t^{\prime} \leq 0.5 t$).
The strength for the NNN hopping is given by the ratio $t^{\prime}/t$, while the adiabaticity ratio is $\omega_{0}/t$.
Throughout this work, unless otherwise indicated, we consider $\omega_{0}/t=1$, and $\lambda_{D} = 0.25$ (or $ g^2/\omega_{0}^2 = 2$), while varying $t^{\prime}/t$ and temperature, $T/t$.

The charge-charge correlations are investigated by the charge structure factor,
\begin{align}\label{eq:Structure_Factor}
S_{\rm cdw}(\textbf{q})  = \frac{1}{L^2} \sum_{i,j} e^{{\rm i}(\textbf{r}_{i}-\textbf{r}_{j})\cdot \textbf{q}}\langle n_{\mathbf{i}} n_{\mathbf{j}} \rangle,
\end{align}
with $L$ being the linear size of the system, i.e., the number of sites being $N=L\times L$.
The peak of $S_{\rm cdw}(\textbf{q})$ defines the leading wavevector \textbf{q} for the charge-charge correlations.
In our case, due to the FSN at half-filling, with $\mathbf{q}_{\rm FSN}=(\pi,\pi)$, we expect to obtain a staggered CDW phase for $t^{\prime}=0$.
In order to probe the CDW critical points, we define the correlation ratio,
\begin{align}\label{eq:Correlation_Ratio}
R_{c}  = 1 - \frac{S_{\rm cdw}(\textbf{Q}-\delta \textbf{q})}{S_{\rm cdw}(\textbf{Q})} ,
\end{align}
in which $\textbf{Q}=(\pi,\pi)$, and $|\delta \textbf{q}|=\frac{2\pi}{L}$.
This quantity exhibits $R_{c} \to 1$ for a long-range ordered phase, and $R_{c} \to 0$ in absence of it.
The critical temperature is obtained from the extrapolation of crossing points of $R_{c}(L)$ curves for different lattice sizes\,\cite{Weber18}.

We also examine the occurrence of metal-insulator transitions from transport properties.
In particular, we analyze the electronic compressibility
\begin{align}\label{eq:Electronic Compressibility}
\kappa = \frac{1}{{\rho}^2}\frac{\partial \rho}{\partial \mu},
\end{align}
with $\rho=\frac{1}{N}\langle\sum_{\mathbf{i}, \sigma} n_{\mathbf{i}, \sigma} \rangle$ being the average electronic density.

We analyze the Hamiltonian of Eq.\,\eqref{eq:Holst_hamil} by the determinant Quantum Monte Carlo (DQMC) method~\cite{Blankenbecler81,Hirsch83,Hirsch85,Scalettar89}
It is an unbiased finite temperature approach in which the non-commuting terms of the Hamiltonian in the partition function are decoupled by performing the Trotter-Suzuki decomposition.
It adds an imaginary-time coordinate, with linear size given by the discretization of the inverse of temperature $\beta=M \Delta\tau$.
Such decomposition leads to an error proportional to $(\Delta\tau)^{2}$, therefore we choose $\Delta\tau=0.1$ in this work, which is enough to systematic errors be smaller than those statistical ones -- from the Monte Carlo sampling.
Furthermore, one may show that, in the Holstein model, the infamous minus-sign problem is absent for any filling, temperature, or interaction strength.

In details, given the Hamiltonian
$$ \mathcal{H} = \mathcal{H}_{K} + \mathcal{H}_{\rm ph} + \mathcal{H}_{\rm el-ph}~, $$
with the right-hand side corresponding to the kinetic, bare phonon modes, and electron-phonon coupling terms, respectively, then one must obtain the partition function
$\mathcal{Z}= \mathrm{Tr}\,e^{-\beta\widehat{\mathcal{H}}}$.
The Trotter-Suzuki transformation leads to
$\mathcal{Z}
\approx
\mathrm{Tr}\,
[\cdots e^{-\Delta\tau\widehat{\mathcal{H}}_{K}} e^{-\Delta\tau\widehat{\mathcal{H}}_{ph}} e^{-\Delta\tau\widehat{\mathcal{H}}_{\rm
el-ph}} \cdots]$, being exact for $\Delta\tau \to 0$.
Here, the trace `Tr' should be performed over the bosonic and fermionic degrees of freedom, leading to
\begin{align}\label{eq:partition_func_Hols}
{\cal Z}  = \int \mathrm{d}\{x_{{\bf i},l}\} & \,e^{-\Delta \tau S_{B}} \times 
\nonumber \\
& \Pi_{\sigma} \bigg[ \mathrm{det}\big(I + B^{\sigma}_{M} B^{\sigma}_{M-1} \cdots B^{\sigma}_{1} \big) \bigg],
\end{align}
in which the matrices $B^{\sigma}_{l}$ are a product of an exponential of the kinetic term and site-diagonal matrices with the exponential of the electron-phonon term, at a given imaginary time slice $l$.
Besides the product of determinants, our statistical weights also have the exponential $e^{-\Delta \tau S_{B}}$, with
\begin{align}\label{eq:phonon_action}
S_{B} = \frac{\omega_{0}^2}{2} \sum_{{\bf i}}
 \sum_{l=1}^{M} \bigg[ 
\frac{1}{\omega_{0}^2 \Delta\tau^2} \big( x_{{\bf i},l} - x_{{\bf i},l+1} \big)^2 
+  x^{2}_{{\bf i},l}  \bigg] ~,
\end{align}
being the bare phonon action, and $\{x_{{\bf i},l}\}$ the set of auxiliary (phonon) fields in real and imaginary-time coordinates.
The integral $\int \mathrm{d}\{x_{i,l}\}$ (i.e., the bosonic trace) is performed by means of Monte Carlo methods.

\begin{figure}[t]
\centering
\includegraphics[scale=0.28]{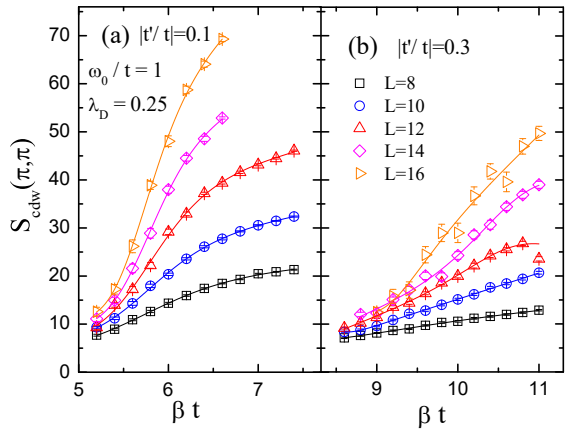}
\caption{Staggered CDW structure factor, $S_{\rm cdw}(\pi,\pi)$, as a function of the inverse temperature for (a) $t^{\prime}/t = -0.1$, and (b) $t^{\prime}/t = -0.3$. The solid lines are just guides to the eye.}
\label{Fig:CDW1}
\end{figure}

A long-standing problem, and one of the bottlenecks in dealing with electron-phonon Hamiltonians by QMC approaches, is the slow phonon dynamics.
Due to the $(\omega_{0}\Delta\tau)^{-2}$ coefficient in the exponential term of the bare phonon action, these systems exhibit long autocorrelation times for either small Trotter discretizations ($\Delta\tau$) or small phonon frequencies $(\omega_{0})$, restricting our QMC analysis.
Over the past years, there were many attempts to overcome this problem\,\cite{Karakuzu18,Chen18,Li19c,Yao21}; here we add our contribution to these discussions by presenting and using another approach.
We perform an \textit{inversion sampling} Monte Carlo method for single moves: it is a no-rejection method in which the changes in a given variable are obtained by inverting the statistical weight~\cite{Fehske07,becca17}.
In the context of phonons, this approach was first developed in Ref.\onlinecite{Costa20}, by two of the authors, for complex auxiliary fields.
In this work, we derive the algorithm for real auxiliary fields.
All the details are presented in the Appendix.
In addition, to ensure that the autocorrelation times are small, we also implement global moves\,\cite{Scalettar91,Johnston13}.


\section{Results} \label{Results}

At this point, we have to mention that the results for $t^{\prime} \neq 0$ have strong finite-size effects.
The inclusion of a NNN hopping term changes the noninteracting Fermi surface, which could lead to open/closed-shell problems at small lattices.
To overcome this problem, we average the quantities over periodic and antiperiodic boundary conditions, for large lattice sizes.
In addition, the following results are restricted to $0 \leq |t^{\prime}/t| \leq 0.5$, a range from which the electronic bandwidth is constant, $W=8t$.

\begin{figure}[t]
\centering
\includegraphics[scale=0.28]{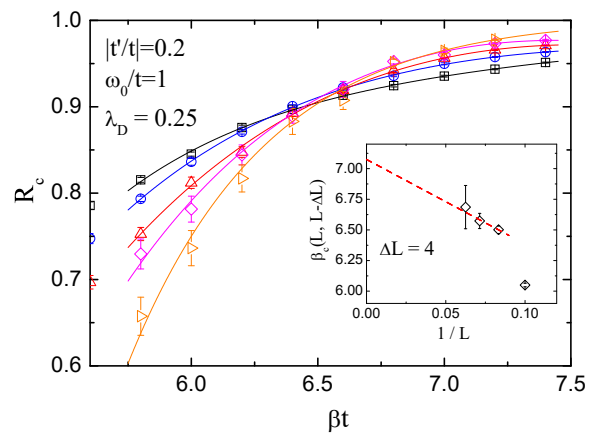}
\caption{Correlation ratio $R_{c}$ as a function of the inverse temperature for $t^{\prime}/t = -0.2$, and different system sizes. The solid lines are just guides to the eye. Inset: the crossing points of $R_{c}(L)$ and $R_{c}(L-\Delta L)$, and their extrapolation to the thermodynamic limit (dashed red line).}
\label{Fig:Rc}
\end{figure}

\begin{figure}[t]
\centering
\includegraphics[scale=0.31]{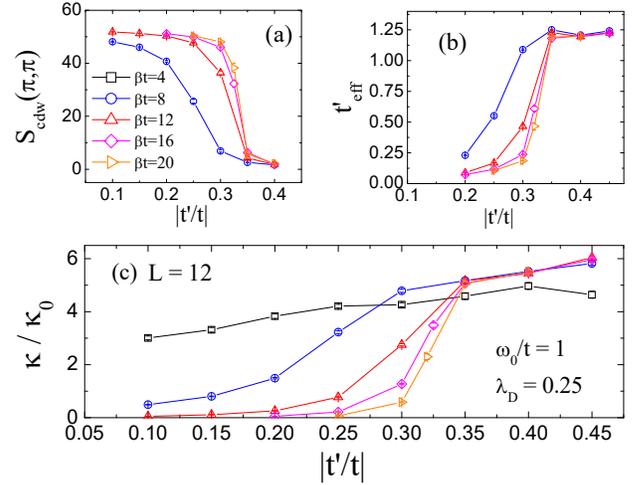}
\caption{(a) Staggered CDW structure factor $S_{\rm cdw}(\pi,\pi)$, (b) the effective next-nearest neighbor hopping $t^{\prime}_{\rm eff}$, and (c) the electronic compressibility $\kappa$ as function of $|t^{\prime}/t|$. Here, we defne $\kappa_{0}(t^{\prime})$ as the ground state compressibility for the noninteracting case, i.e.~$g=0$.}
\label{Fig:Compress}
\end{figure}

We start discussing the effects of $t^{\prime}$ on the charge-charge correlations.
For fixed $t^{\prime} = 0$, $\omega_{0}/t=1$, and $\lambda_{D} = 0.25$, the Holstein model exhibits CDW long-range order below the critical temperature $T_{c}= 0.174(2)$\,\cite{Weber18,Batrouni19}.
That is, for $\beta t \gtrsim 5.75$, the charge structure factor should increase with the lattice size, diverging at the thermodynamic limit.
Interestingly, the behavior for $t^{\prime}/t = -0.1$ is similar, as displayed in Fig.\,\ref{Fig:CDW1}\,(a), in which $S_{\rm cdw}(\pi,\pi)$ has a strong dependence with the lattice size for $\beta t \gtrsim 5.8$ or 6.0.
Therefore, one may expect CDW order for $t^{\prime}/t = -0.1$ around this temperature scale.
The saturation of $S_{\rm cdw}(\pi,\pi)$ at higher $\beta t$ (lower temperatures) is a finite size effect, indicating that the charge correlation length is larger than the linear dimension of the lattice, i.e. $\xi/L \gtrsim 1$.
However, increasing the NNN hopping leads to quite different results.
For instance, at $t^{\prime}/t = -0.3$, the charge structure factor is supressed at higher temperatures, and its strong dependence with the lattice size is noticed only for $\beta t \gtrsim 9.5$, as displayed in Fig.\,\ref{Fig:CDW1}\,(b).
Further increase in $t^{\prime}$ suppresses $S_{\rm cdw}(\pi,\pi)$ even at very low temperatures, as $\beta t \approx 20$, for any lattice size [see, e.g., Fig.\,\ref{Fig:Compress}\,(a)].

In order to identify the occurrence of long-range order, we investigate the behavior of the correlation ratio, Eq.\,\eqref{eq:Correlation_Ratio}.
For instance, Fig.\,\ref{Fig:Rc} displays $R_{c}$ as a function of the inverse of temperature, for different lattice sizes, at fixed $t^{\prime}/t = -0.2$.
The crossings of the curves around $\beta t \approx 6.6$ indicate that the CDW phase should emerge around this energy scale.
A more precise determination of the critical temperature is performed by extrapolating the crossing points between $R_{c}(L)$ and $R_{c}(L-\Delta L)$ -- defined as $\beta_{c}(L,L-\Delta L)$ -- to $L \to \infty$, as presented in the inset of Fig.\,\ref{Fig:Rc}.
In this case, for $t^{\prime}/t = -0.2$, we have found $\beta_{c} t = 7.1(4)$.

\begin{figure}[t]
\centering
\includegraphics[scale=0.28]{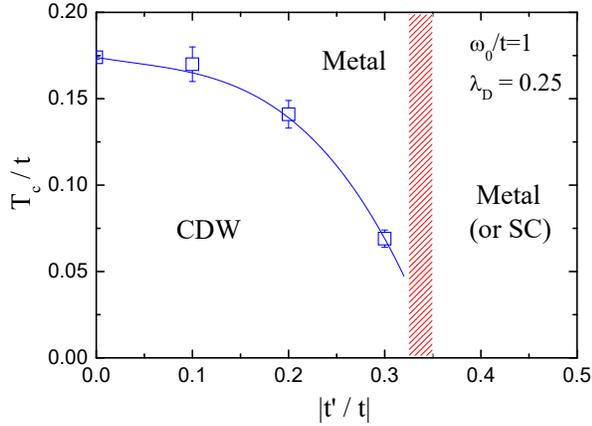}
\caption{Critical temperatures for $\lambda_{D}=0.25$ and $\omega_{0}/t=1$ as function of $|t^{\prime}/t|$. The red bar defines the range of $|t^{\prime}/t|$ at which an abrupt (first-order-like) phase transition occurs at low temperatures.}
\label{Fig:PhaseDiagram}
\end{figure}

Repeating the same procedure for other values of NNN hopping, we obtain the finite-temperature phase diagram displayed in Fig.\,\ref{Fig:PhaseDiagram}. 
One may notice that the critical temperature is reduced as $|t^{\prime}/t|$ increases, and has a strong suppression for $t^{\prime}/t \approx -0.35$.
By examining the behavior of $S_{\rm cdw}(\pi,\pi)$ as a function of $|t^{\prime}/t|$, we notice an abrupt change in its response, as shown in Fig.\,\ref{Fig:Compress}\,(a), consistent with a first-order phase transition.
These results show how harmful is the inclusion of further neighbor hoppings to the occurrence of a staggered CDW phase.

At this point, it is important recalling the relevance of the noninteracting Fermi surface to the emergence of charde order.
For $t^{\prime}=0$, the existence of a FSN and a van Hove singularity at the half-filled square lattice lead to a logarithm divergence in the electronic susceptibility, $\chi_{0} \sim \ln^{2}(t/T)$.
Therefore, the system may exhibit charge instabilities in presence of any electron-phonon coupling, as suggested by recent QMC results\cite{Hohenadler19,Costa20}.
Interestingly, even when the van Hove singularity is destroyed -- but maintaining the FSN --, the half-filled square lattice still has a stable CDW phase as a function of external parameters, as discussed in Ref.\,\onlinecite{CohenStead19} for fixed $\lambda_{D}=0.25$.
However, the inclusion of a NNN hopping destroys both FSN and van Hove singularity, in particular at the weak electron-phonon coupling.
That is, despite being important, the FSN is not the key ingredient to understand the emergence, and eventually, the suppression of the CDW phase in our case, in particular at the intermediate coupling strength $\lambda_{D}=0.25$~\footnote{For instance, this interaction strength is enough to lead to a CDW phase in the half-filled honeycomb lattice, as presented in Refs.\onlinecite{Zhang19,Costa21}.}.

To shed light on it, we have to discuss the transport properties of the model.
We start investigating the effective NNN hopping~\cite{Varney09}
\begin{align}
t^{\prime}_{\rm eff} =
\frac{\langle d^{\dagger}_{\mathbf{i} \sigma} d^{\phantom{\dagger}}_{\mathbf{j} \sigma} + d^{\dagger}_{\mathbf{j} \sigma} d^{\phantom{\dagger}}_{\mathbf{i} \sigma} \rangle_{\lambda_{D}}}
{\langle d^{\dagger}_{\mathbf{i} \sigma} d^{\phantom{\dagger}}_{\mathbf{j} \sigma} + d^{\dagger}_{\mathbf{j} \sigma} d^{\phantom{\dagger}}_{\mathbf{i} \sigma} \rangle_{\lambda_{D} = 0}}~,
\end{align}
with $\mathbf{j} = \mathbf{i} \pm \hat{x} \pm \hat{y}$.
Figure \ref{Fig:Compress}\,(b) presents the effective hopping as a function of $t^{\prime}$, for different temperatures.
Notice that, for $|t^{\prime}/t| \lesssim 0.35$, the tendency is for the suppression of $t^{\prime}_{\rm eff}$, i.e.~the hopping between NNN sites is not allowed, due to the staggered double occupancy distribution of electrons.
For larger $t^{\prime}$, the effective hopping has a sharp increase, while $S_{\rm cdw}(\pi,\pi)$ is suppressed.
Such a behavior, could reflect the competition between staggered and striped CDW phases, since the system could also reduce its energy by having a striped order.
However, our results for $\lambda_{D}=0.25$ and $t^{\prime}/t \leq 0.5$ provide no enhancement of $(0,\pi)$ charge-charge correlations.
That is, the enhancement in $t^{\prime}_{\rm eff}$ (for $|t^{\prime}/t| \approx 0.35$), and the corresponding suppression of $S_{\rm cdw}(\pi,\pi)$ should have their nature in charge frustration effects\,\cite{Li18}.
\footnote{It is analogous to the suppression of magnetism in the $J_{1}-J_{2}$ Heisenberg model -- see, e.g., Refs.\,\onlinecite{Zhong93,Cysne15,Ferrari20}.}

\begin{figure}[t]
\centering
\includegraphics[scale=0.28]{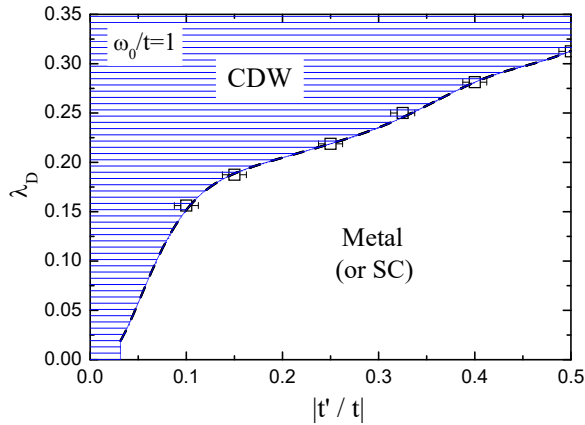}
\caption{Low-temperature phase diagram for the two-dimensional $t-t^{\prime}$ Holstein model for fixed $\omega_{0}/t=1$. The data points were obtained for $\beta t=20$ and $L=12$. The solid line is just a guide to the eye.}
\label{Fig:MIT_Phase}
\end{figure}

The suppression of long-range charge-charge correlations leads to a metallic or superconducting ground state.
Therefore we finish our analysis by examining the occurrence of metal-to-insulator transitions at low temperatures.
We investigate the behavior of the electronic compressibility as a function of $t^{\prime}$, as presented in Fig.\,\ref{Fig:Compress}\,(c), for different temperatures.
Here, $\kappa_{0}$ denotes the ground state compressibility for the noninteracting case (and also for its corresponding $t^{\prime}$), at the thermodynamic limit.
Similar to the effective NNN hopping term, $\kappa$ has a sharp change around $|t^{\prime}/t| \approx 0.35$ for $\beta t =20$, determining the occurrence of a metal-to-insulator transition at this point.
Indeed, the examination for higher temperatures shows that $\kappa$ has different tendencies below and above such a critical point.
Repeating the same procedure to other values of $\lambda_{D}$, with fixed $\omega_{0}/t = 1$, also combined with the analysis of the $S_{\rm cdw}(\pi,\pi)$ behavior, we obtain the low temperature phase diagram presented in Fig.\,\ref{Fig:MIT_Phase}, which displays the emergence of a CDW phase for different values of $\lambda_{D}$ and $t^{\prime}$.

It is important to give some remarks about this low-temperature phase diagram.
First, concerning the occurrence of superconductivity, we have measured the $s$-wave superconducting pair susceptibility, $\chi_{s} = \frac{1}{N} \int^{\beta}_{0} \mathrm{d}\tau \langle \Delta(\tau) \Delta^{\dagger}(0)\rangle$, with $\Delta(\tau) = \sum_{\mathbf{i}} c_{\mathbf{i}\downarrow}(\tau) c_{\mathbf{i}\uparrow}(\tau)$ at the metallic side of the phase diagram.
For $\beta t \leq 20$, although $\chi_{s}$ increases when the temperature is reduced, it presents just a weak dependence with the lattice size (not shown).  
As already pointed out in the literature, the critical temperature for the Kosterlitz-Thouless transition at fixed $\omega_{0}/t = 1$ should be very low\,\cite{Vekic92,Dee19,Bradley21}, and unfeasible to reach from a technical point of view.
However, due to the suppression of the charge-charge correlations, the attractive effective interaction between electrons must lead to a SC phase at the ground state\,\cite{Huscroft97,Costa18,Xiao21b}, for $|t^{\prime}/t| \geq 0.35$. 

As a second remark, we also noticed that, for larger $\lambda_{D}$, the metal-to-insulator transition becomes even more abrupt, in line with a first-order phase transition.
We believe that it occurs due to a residual competition between CDW and SC correlations at high temperatures.
Since these two phases break different symmetries, a ground state phase transition between them should be first-order-like.
Interestingly, our results show that, even in absence of SC at high temperatures, the competition between these tendencies is enough to lead to a first-order metal-to-insulator transition.  
As a final remark, for the range of parameters we analyzed, $0 < |t^{\prime}/t| \leq 0.50$, and $1.25 t \leq W \lambda_{D} \leq 2.50 t$, we have not found CDW striped phases, but we expect that these phases would appear for larger values of $t^{\prime}/t$.

\section{Conclusions} \label{Conclusions}

In this work, we have examined the properties of the $t$-$t^{\prime}$ Holstein model in the half-filled square lattice, by QMC simulations.
On the methodological side, we optimized the Monte Carlo sampling by implementing an inversion sampling algorithm, as described in Section \ref{Sec:ModelandMethod}, and in the Appendix.
This approach strongly reduces the autocorrelation time, a problem that affects the analysis of electron-phonon Hamiltonians.
In principle, this kind of sampling could be adapted to other models, such as the Hubbard model with continuous Hubbard-Stratonovich fields.

On the physical side, we analyzed the occurence of long-range order by means of the charge structrure factor, and its correlation ratio, by fixing $\omega_{0}/t=1$.
For $\lambda_{D}=0.25$, which is an intermediate interaction strength value, we determined the occurrence of staggered CDW phase transitions at finite critical temperatures $T_{c}$, for any $\mid t^{\prime}/t\mid < 0.35 $.
However, frustration effects are harmful to this charge-ordered phase and, for $\mid t^{\prime}/t\mid \gtrsim 0.35$, there is a strong suppression of the charge-charge correlations, leading to a metallic (or superconducting) phase.
This metal-to-insulator transition seems to be consistent with a first-order phase transition, which could be a residual feature of the competition between CDW and SC correlations at low temperatures.
We also provided a phase diagram of the model for intermediate interaction strength values of $\lambda_{D}$.
These results provide a better understanding of the competition between CDW and SC, in particular for the suppression of the former by pressure.
The analysis away from half-filling, and adding electron-electron interactions could be relevant to the cuprate physics, but it is beyond the scope of this work.

\begin{acknowledgements}
This work is supported by the European Centre of Excellence in Exascale Computing TREX - Targeting Real Chemical Accuracy at the Exascale. This project has received funding from the European Union's Horizon 2020 - Research and Innovation program - under grant agreement no.~952165.
Computational resources were provided by CINECA supercomputer, project IsB23 (ISCRA-HP10BF65I0).
N.C.C.~and S.S.~acknowledge ISCRA for awarding them access to Marconi100 at CINECA, Italy.
S.S.~also acknowledge financial support by the MIUR Progetti di Ricerca di Rilevante Interesse Nazionale (PRIN) Bando 2017 - Grant No.~2017BZPKSZ.
M.V.A., J.P.L., and N.C.C.~acknowledge L.~Oliveira-Lima for the discussions at the beggining of this work, and the Brazilian Agencies National Council for Scientific and Technological Development (CNPq), National Council for the Improvement of Higher Education (CAPES), and FAPERJ for partially funding this project.
N.C.C. particularly thanks S.\,Sorella for the discussions during his QMC course in SISSA, from which the methodology developed in this paper arised.  
\end{acknowledgements}

\appendix* 
\section{Inversion sampling Monte Carlo method}
\label{Ap}
In this Appendix, we present how to implement the inversion sampling Monte Carlo algorithm to single site updates in the Holstein model. In fact, the same procedure could be performed for continuous Hubbard-Stratonovich fields, in particular to the attractive Hubbard model.

Here, we start from the fast Green's functions update procedure.
For a given site $\mathbf{i}$, time slice \textit{l}, and spin sector $\sigma$, one may show that the fermionic determinant weight related to a change $ x_{\mathbf{i},l} \to x^{\prime}_{\mathbf{i},l} \equiv x_{\mathbf{i},l} + \delta x_{\mathbf{i},l}$ is
\begin{align}
R^{\sigma} = G^{\sigma}_{\mathbf{i}\mathbf{i}}(l) + [1 - G^{\sigma}_{\mathbf{i}\mathbf{i}}(l)] e^{\Delta\tau g \delta x_{\mathbf{i},l}}~,
\end{align}
with $G^{\sigma}_{\mathbf{i}\mathbf{i}}(l)$ being the equal-time Green's functions.
Due to the product of determinants of different spin sectors, we have
\begin{align}
\nonumber
R^{\uparrow} R^{\downarrow} = & ~G^{\uparrow}_{\mathbf{i}\mathbf{i}}(l) G^{\downarrow}_{\mathbf{i}\mathbf{i}}(l)
+ G^{\uparrow}_{\mathbf{i}\mathbf{i}}(l)[1 - G^{\downarrow}_{\mathbf{i}\mathbf{i}}(l)] e^{\Delta\tau g \delta x_{\mathbf{i},l}}
\\ \nonumber
+ & ~G^{\downarrow}_{\mathbf{i}\mathbf{i}}(l)[1 - G^{\uparrow}_{\mathbf{i}\mathbf{i}}(l)] e^{\Delta\tau g \delta x_{\mathbf{i},l}}
\\
+ & ~[1 - G^{\uparrow}_{\mathbf{i}\mathbf{i}}(l)] [1 - G^{\downarrow}_{\mathbf{i}\mathbf{i}}(l)] e^{2\Delta\tau g \delta x_{\mathbf{i},l}}.
\end{align}
For the particular case of having $G^{\downarrow}_{\mathbf{i}\mathbf{i}}(l) = G^{\uparrow}_{\mathbf{i}\mathbf{i}}(l)$ -- which occurs for the Holstein model --, then we obtain
\begin{align}
R^{\uparrow} R^{\downarrow} = p_{1} + p_{2} e^{\Delta\tau g \delta x_{\mathbf{i},l}}
+ p_{3}  e^{2\Delta\tau g \delta x_{\mathbf{i},l}} ~,
\end{align}
with $p_{1}= G^{\uparrow}_{\mathbf{i}\mathbf{i}}(l)^2$,
$p_{2} = 2 G^{\uparrow}_{\mathbf{i}\mathbf{i}}(l)[1 - G^{\uparrow}_{\mathbf{i}\mathbf{i}}(l)]$, and
$p_{3} = [1 - G^{\uparrow}_{\mathbf{i}\mathbf{i}}(l)]^2$.

As described in Eq.\,\eqref{eq:partition_func_Hols}, the total statistical weight has also a dependency with changes in the bosonic phonon action, i.e. $r = e^{-\Delta\tau (S^{\prime}_{B} - S_{B})} R^{\uparrow} R^{\downarrow} $, with $S^{\prime}_{B}$ being the action for the updated variable.
Using the definition of Eq.\,\eqref{eq:phonon_action}, one may also show that
$$ S^{\prime}_{B} - S_{B} =  A [ \delta x_{\mathbf{i},l}^2 + 2 C_{\mathbf{i},l} \delta x_{\mathbf{i},l}] ~,$$
with $A=\frac{\omega^{2}_{0}}{2} \big( 1 + \frac{2}{\omega^{2}_{0} \Delta\tau^2} \big)$, and
$C_{\mathbf{i},l}= \frac{\langle x_{\mathbf{i}} \rangle_{l}}{2 + \omega^{2}_{0}\Delta\tau^2} - x_{\mathbf{i},l}$, and
with  $\langle x_{\mathbf{i}} \rangle_{l} = \frac{x_{\mathbf{i},l-1} + x_{\mathbf{i},l+1}}{2}$.
Therefore,
\begin{align}
\nonumber
r = & e^{-\Delta\tau A [ \delta x_{\mathbf{i},l}^2 + 2 C_{\mathbf{i},l} \delta x_{\mathbf{i},l}] }
\\
& \times \big[ p_{1} + p_{2} e^{\Delta\tau g \delta x_{\mathbf{i},l}}
+ p_{3}  e^{2\Delta\tau g \delta x_{\mathbf{i},l}} \big].
\end{align}
Notice that, in fact, it is the sum of three Gaussians
\begin{align}\label{Ap:StatWeight}
\nonumber
r = & W_{1} \times e^{-\Delta\tau A [ \delta x_{\mathbf{i},l} - C_{\mathbf{i},l} ]^2 }
\\ \nonumber
+ & W_{2} \times e^{-\Delta\tau A [ \delta x_{\mathbf{i},l} - (C_{\mathbf{i},l} + g/2A) ]^2 }
\\
+ & W_{3} \times e^{-\Delta\tau A [ \delta x_{\mathbf{i},l} - (C_{\mathbf{i},l} + g/A) ]^2 },
\end{align}
with
\begin{align*}
& W_{1} = p_{1} e^{\Delta\tau A C_{\mathbf{i},l}^2}~,  \\
& W_{2} = p_{2} e^{\Delta\tau A (C_{\mathbf{i},l} + g/2A)^2}~, \\
& W_{3} = p_{3} e^{\Delta\tau A (C_{\mathbf{i},l} + g/A)^2}~.
\end{align*}

At this point, it is important to normalize the statistical weight of Eq.\,\eqref{Ap:StatWeight}.
Since
\begin{align}
\int_{-\infty}^{\infty} r(\delta x_{\mathbf{i},l}) ~ {\rm d}[\delta x_{\mathbf{i},l}] =
\sqrt{\frac{\pi}{\Delta\tau A}} \big( W_{1} + W_{2} + W_{3} \big),
\end{align}
then the normalized distribution becomes
\begin{align}\label{3gauss}
\nonumber
\tilde{r} = & \tilde{W}_{1} \times  e^{-\Delta\tau A [ \delta x_{\mathbf{i},l} - C_{\mathbf{i},l} ]^2 } \\
\nonumber
+ & \tilde{W}_{2} \times  e^{-\Delta\tau A [ \delta x_{\mathbf{i},l} - (C_{\mathbf{i},l} + g/2A) ]^2 } \\
+ & \tilde{W}_{3} \times  e^{-\Delta\tau A [ \delta x_{\mathbf{i},l} - (C_{\mathbf{i},l} + g/A) ]^2 },
\end{align}
with
\begin{align*}
\tilde{W}_{k} = \frac{\sqrt{\frac{\Delta\tau A}{\pi}} W_{k}}{\sum_{j} W_{j}}.
\end{align*}
Now, this normalized probability distribution of three Gaussians can be sampled exactly by the Box-M\"uller method.
That is, we obtain $\delta x_{\mathbf{i},l}$ by inverting the distribution of Eq.\,\eqref{3gauss}, which leads to a \textit{no-rejection} sampling. For more details, we recommend the pedagogical discussions at Ref.\onlinecite{becca17}.


\bibliography{bibAraujo_Hols}

\end{document}